\documentclass[usenatbib,letterpaper]{mn2e}
\usepackage[totalwidth=480pt,totalheight=680pt,twoside=false,voffset=.5cm]{geometry}

\usepackage{graphics,graphicx,epsfig,ulem,multirow,appendix,lscape,amsmath,amssymb,color,url,float}

\long\def\symbolfootnote[#1]#2{\begingroup%
\def\thefootnote{\fnsymbol{footnote}}\footnote[#1]{#2}\endgroup} 

\begin{document}

\title[Equal-mass planet disc interactions]
  {Double-ringed debris discs could be the work of eccentric planets: explaining the strange morphology of HD 107146}
\author[T. D. Pearce \& M. C. Wyatt]
  {Tim D. Pearce\thanks{tdpearce@ast.cam.ac.uk} and Mark C. Wyatt\\
  Institute of Astronomy, University of Cambridge, Madingley Road, Cambridge, CB3 0HA, UK}
\date{Released 2002 Xxxxx XX}

\pagerange{\pageref{firstpage}--\pageref{lastpage}} \pubyear{2002}

\def\LaTeX{L\kern-.36em\raise.3ex\hbox{a}\kern-.15em
    T\kern-.1667em\lower.7ex\hbox{E}\kern-.125emX}

\newtheorem{theorem}{Theorem}[section]

\label{firstpage}

\maketitle


\begin{abstract}              
\noindent We investigate the general interaction between an eccentric planet and a coplanar debris disc of the same mass, using analytical theory and $n$-body simulations. Such an interaction could result from a planet-planet scattering or merging event. We show that when the planet mass is comparable to that of the disc, the former is often circularised with little change to its semimajor axis. The secular effect of such a planet can cause debris to apsidally \textit{anti-}align with the planet's orbit (the opposite of what may be na\"{i}vely expected), leading to the counter-intuitive result that a low-mass planet may clear a larger region of debris than a higher-mass body would. The interaction generally results in a double-ringed debris disc, which is comparable to those observed in HD 107146 and HD 92945. As an example we apply our results to HD 107146, and show that the disc's morphology and surface brightness profile can be well-reproduced if the disc is interacting with an eccentric planet of comparable mass ($\sim 10-100$ Earth masses). This hypothetical planet had a pre-interaction semimajor axis of 30 or 40 au (similar to its present-day value) and an eccentricity of 0.4 or 0.5 (which would since have reduced to $\sim 0.1$). Thus the planet (if it exists) presently resides near the inner edge of the disc, rather than between the two debris peaks as may otherwise be expected. Finally we show that disc self-gravity can be important in this mass regime and, whilst it would not affect these results significantly, it should be considered when probing the interaction between a debris disc and a planet.
\end{abstract}



\section{Introduction}
\label{sec: Introduction}

\noindent Given the calm order of the Solar System today, where most planets and minor bodies occupy near circular and coplanar orbits, one could be forgiven for forgetting that planetary systems can be violent places. Indeed, our own system probably had a tumultuous youth; planets may have scattered off each other and collided (\citealt{Hartmann75}), switched places with one-another (\citealt{Tsiganis05}) and ploughed into regions of debris (\citealt{Walsh11}). This turbulent picture is also inferred from extrasolar planets; many of these objects are eccentric \citep{Schneider11}, a possible hallmark of previous scattering events \citep{Juric08}. Some, such as Fomalhaut b, may also pass through regions of debris \citep{Kalas13, Pearce15}. Furthermore, major orbital evolution is required to explain some classes of extrasolar planets, such as the Hot Jupiters (\citealt{Rasio96, Wu11}). So if planet-planet scattering, mergers and dynamical instabilities could be the norm then it is pertinent to ask how planets affected by these processes interact with other bodies in the system, and whether we can use this information to probe their past or ongoing dynamical evolution.

In this paper we examine the general evolution of a system hosting a debris disc interacting with an equal-mass, coplanar, eccentric planet, assuming the planet's eccentricity was rapidly driven up by one of the above processes. Our Solar System hosts two debris discs, the Asteroid and Kuiper Belts, and many extrasolar discs have been detected as infrared excesses in the spectra of stars (e.g. \citealt{Rhee07, Eiroa13}). So given that debris discs are reasonably common, it is likely that dynamically evolving planets interact with these structures in at least some systems. Instruments such as Spitzer, the HST and ALMA have resolved some extrasolar debris discs (e.g. \citealt{Backman09, Schneider09, Dent14}), and the eccentricity or clumpiness of these discs can be used to infer the presence of planets which would be otherwise undetectable. Hence one aim of this work is to characterise this interaction in general, to ascertain the signatures an eccentric planet leaves on a disc which can then be compared to observations. This part of the paper is similar in its goals and methodology to our previous investigation \citep{Pearce14}, in which we considered only planets much more massive than the disc; the main difference now that the planet and disc are of equal mass is the disc's ability to significantly alter the planet's orbit, which can have major effects on the system evolution.

Our second aim is to apply the general results to the debris disc of HD 107146, which has been resolved in both infrared emission and scattered light \citep{Ardila04, Williams04, Carpenter05, Corder09, Hughes11, Ricci15}. This disc is seen nearly face on, and is broad ($\sim 100$ au) and axisymmetric with a $\sim 30$ au inner hole. The curiosity of this system is that the 1.25mm debris surface density profile appears to either first decrease and then increase with radius, or generally increase but with a gap at around 80 au \citep{Ricci15}. This differs from the surface density profiles of protoplanetary discs, which decrease with radius (e.g \citealt{Andrews07}); it has been suggested that the unusual profile could be the result of planetary perturbations. Given the large disc mass (possibly $\sim 100 {\rm M}_\oplus$ in total; \citealt{Ricci15}), this system is a prime application of our results. We will show that the strange disc morphology can be explained as the aftermath of the interaction between an eccentric planet and a coplanar debris disc of the same mass.

The layout of this paper is as follows. We examine the general outcomes of this interaction in Section \ref{sec: general_results}, using a combination of theory and $n$-body simulations. We then apply these results to HD 107146 in Section \ref{sec: HD107146 Application}, where we run simulations over a broad region of parameter space to replicate the disc structure and surface brightness profile observed with ALMA. We discuss the implications of the results in Section \ref{sec: discussion}, and conclude in Section \ref{sec: conclusions}.


\section{General interaction outcomes}
\label{sec: general_results}

\noindent In this section we describe the general outcomes of the interaction between an eccentric planet and a comparable-mass, coplanar debris disc. We first summarise the dynamical processes involved in the interaction in sections \ref{sec: secular_interactions} to \ref{sec: planet_evolution}, and describe their effects on the orbits of both the debris particles and the planet. Wherever possible we give quantitative predictions from dynamical theory. In Section \ref{sec: simulations} we present an $n$-body simulation as an example of the general outcomes of this interaction, and explain these in the context of the theoretical results.

\subsection{Secular effects on debris}
\label{sec: secular_interactions}

\begin{figure*}
  \centering
   \includegraphics[width=16cm]{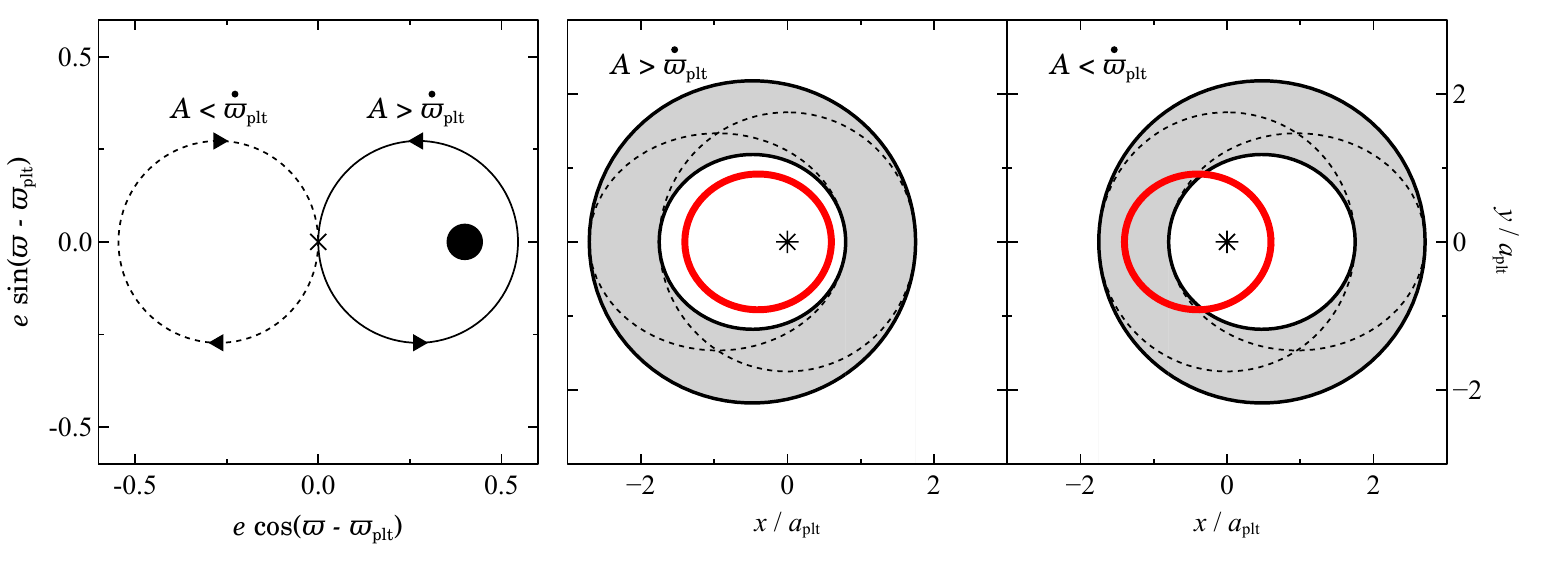}
   \caption{Evolution of a test particle's orbit under the influence of a single, precessing planet, according to second-order secular theory. Shown are the cases where its orbit precesses more quickly than that of the planet ($A > \dot{\varpi}_{\rm plt}$) and more slowly ($A < \dot{\varpi}_{\rm plt}$), for a planet of a given eccentricity. Left plot: the coupled evolution of eccentricity $e$ and longitude of pericentre $\varpi$, where the large black circle denotes the planet. A particle of a given semimajor axis will move around one of the two paths in the direction indicated, depending on the sign of $A - \dot{\varpi}_{\rm plt}$. Right plots: the corresponding physical areas swept out by the particles, in a frame instantaneously aligned with the planet's orbit. The asterisk and thick red line denote the star and the planet's orbit respectively, and the dashed lines show the extreme orbits of the particle. They grey region is the area swept out by the particle, resulting from the superposition of all orbits between the two extremes. Both plots show systems with identical parameters, but with differing signs of $A - \dot{\varpi}_{\rm plt}$. Hence a particle which would never cross the planet's orbit if $A > \dot{\varpi}_{\rm plt}$ may do so if $A < \dot{\varpi}_{\rm plt}$. Also note that the planet is precessing, so the structures on the right hand plots will rotate whilst remaining aligned with the planet's orbit.}
   \label{fig: ew_evo}
\end{figure*}

\noindent Secular interactions are long-term angular momentum exchanges between bodies, which can cause a particle's eccentricity and orbital plane (but not semimajor axis) to evolve. As secular timescales are much longer than orbital timescales then, providing the objects are not in a mean-motion resonance, each interacting body may be thought of as an extended ``wire'' of material in the shape of the object's orbit, with a density at each point inversely proportional to the body's velocity there (this approach to secular perturbations is known as Gauss averaging). The influence of other masses causes this wire to change shape and orientation. There exists no general analytic evaluation of this interaction (although semi-analytic solutions can be constructed in some cases; see \citealt{Beust14}), so a common approach is to derive an analytical form complete up to second order in eccentricities and inclinations, and disregard all higher terms. This is valid for small eccentricities and inclinations, but introduces errors if larger values are considered; however, we showed in \citet{Pearce14} that this still produces qualitatively correct results for very large eccentricities so long as the inclinations are small. We now summarise the behaviour of a particle undergoing a secular interaction with a planet according to second-order secular theory, and apply the results to the interaction between an eccentric planet and a debris disc.

Second-order secular theory predicts that a body's eccentricity $e$ and longitude of pericentre $\varpi$ are coupled (for details see \citealt{Murray_Dermott99}). Specifically, for a test particle undergoing secular perturbations from one or more massive bodies, these quantities satisfy the equations

\begin{align}
 e \cos \varpi &= e_{\rm p} \cos \varpi_{\rm p} + e_{\rm f} \cos \varpi_{\rm f}, \label{eq: ecosw}\\
 e \sin \varpi &= e_{\rm p} \sin \varpi_{\rm p} + e_{\rm f} \sin \varpi_{\rm f}. \label{eq: esinw}
\end{align}

\noindent Here $e_{\rm p}$ and $\varpi_{\rm p}$ denote the ``free'' parameters of the particle; $e_{\rm p}$ is constant, and $\varpi_{\rm p}$ increases linearly with time $t$ such that

\begin{equation}
\varpi_{\rm p} = A t + \beta,
\label{eq: wp}
\end{equation}

\noindent where $A$ and $\beta$ are constants. The quantities $e_{\rm f}$ and $\varpi_{\rm f}$ are the ``forced'' values, which depend on the parameters of the massive bodies in the system and may evolve in time. Hence the test particle moves around a circle on the $e \cos \varpi$, $e \sin \varpi$ plane, of radius $e_{\rm p}$ and at a rate $A$. Meanwhile the centre of this circle also moves as the perturbing bodies evolve over time.

We now consider a system comprised of a star of mass $M_*$, a test particle at semimajor axis $a$ and a planet of mass $M_{\rm plt}$ at semimajor axis $a_{\rm plt}$ (where $a > a_{\rm plt}$). We also assume that some mechanism causes the planet's pericentre to precess ($\dot{\varpi}_{\rm plt} \neq 0$). We set both the test particle's eccentricity and the planet's longitude of pericentre to zero at time zero for simplicity, i.e. $\beta = \pi$, $\varpi_{\rm f} = \varpi_{\rm plt}$, $e_{\rm p} = |e_{\rm f0}|$ (where $e_{\rm f0}$ is $e_{\rm f}$ evaluated at $t=0$). The forcing eccentricity is now

\begin{equation}
e_{\rm f} = \frac{|A_{\rm plt}|}{A - \dot{\varpi}_{\rm plt}}e_{\rm plt},
\label{eq: e_forced}
\end{equation}

\noindent where

\begin{align}
A_{\rm plt} = -\frac{1}{4} \sqrt{\frac{G M_*}{a^3}} \frac{M_{\rm plt}}{M_*} \frac{a_{\rm plt}}{a} b^{(2)}_{3/2}(a_{\rm plt}/a), \label{eq: Aplt}\\
A = \frac{1}{4} \sqrt{\frac{G M_*}{a^3}} \frac{M_{\rm plt}}{M_*} \frac{a_{\rm plt}}{a} b^{(1)}_{3/2}(a_{\rm plt}/a), \label{eq: A}
\end{align}

\noindent $G$ is the gravitational constant and $b^{(j)}_{s}(\alpha)$ is a Laplace coefficient, such that

\begin{equation}
b^{(j)}_{s}(\alpha) \equiv \frac{1}{\pi} \int^{2\pi}_0 \frac{\cos(j\psi){\rm d}\psi}{(1-2\alpha \cos \psi + \alpha^2)^s}.
\label{eq: laplace_coeff}
\end{equation}

Equation \ref{eq: e_forced} shows that the orbit of the test particle will evolve differently in the regimes $A > \dot{\varpi}_{\rm plt}$ and $A < \dot{\varpi}_{\rm plt}$. Firstly, if $A > \dot{\varpi}_{\rm plt}$ then the precession rate of the particle is faster than that of the planet, and $e_{\rm f} > 0$. The particle will therefore move anticlockwise about a circle on the $ e \cos(\varpi - \varpi_{\rm plt})$, $e \sin(\varpi - \varpi_{\rm plt})$ plane, which crosses the origin and is offset from the origin in the direction of the planet. This is shown by the solid line on the left hand plot of Figure \ref{fig: ew_evo}. Hence the particle's eccentricity will be maximised when its orbit is aligned with that of the planet, and small when antialigned. Superimposing all intermediate orbits shows that the particle will sweep out a broad, eccentric disc aligned with the planet's orbit (central plot of Figure \ref{fig: ew_evo}). This allows the particle to attain high eccentricities and yet be shielded from scattering by the planet. For a complete summary of particle evolution in this regime, see \citet{Pearce14} and \citet{Faramaz14}.

If $A < \dot{\varpi}_{\rm plt}$, the planet precesses more quickly than the particle and the forcing eccentricity becomes negative. The particle will now move clockwise about a circle on the $ e \cos(\varpi - \varpi_{\rm plt})$, $e \sin(\varpi - \varpi_{\rm plt})$ plane, again crossing the origin but offset in the direction opposing the planet. This is the dashed line on the left hand plot of Figure \ref{fig: ew_evo}. The particle's eccentricity will now be maximised when its orbit is antialigned with that of the planet, and superimposing all intermediate orbits results in a broad, eccentric disc antialigned with the planet's orbit (right hand plot of Figure \ref{fig: ew_evo}). Hence a particle which would never cross the planet's orbit if $A > \dot{\varpi}_{\rm plt}$ may now do so, and could therefore be ejected by the latter.

\begin{figure}
  \centering
   \includegraphics[width=7cm]{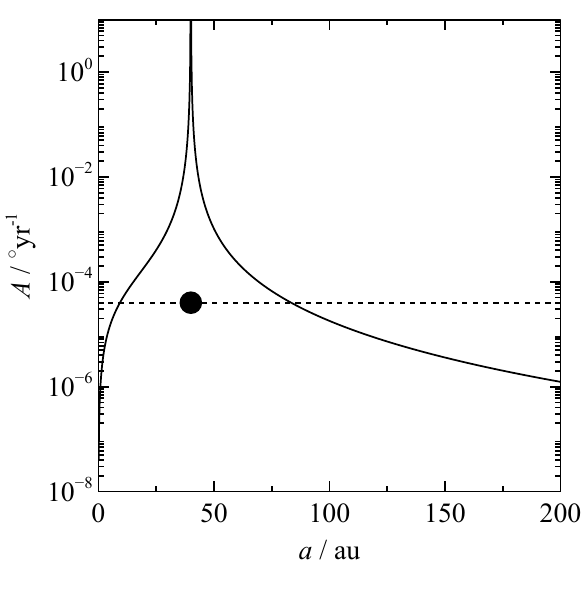}
   \caption{Precession rate of a test particle as a function of its semimajor axis, derived using second-order secular theory. Here the planet (black circle) has a semimajor axis of 40 au and precesses at a rate $\dot{\varpi}_{\rm plt} = 4 \times 10^{-5}$ $^\circ {\rm yr}^{-1}$. Particles precessing more rapidly than the planet form an eccentric disc apsidally aligned with the planet's orbit (central plot on Figure \ref{fig: ew_evo}), whilst those with slower precession rates will antialign with the planet's orbit (right hand plot on Figure \ref{fig: ew_evo}). Note that this plot does not take account of mean motion resonances, the effect of which could dominate over secular behaviour.}
   \label{fig: A_vs_a}
\end{figure}

\begin{figure}
  \centering
   \includegraphics[width=5.716666666cm]{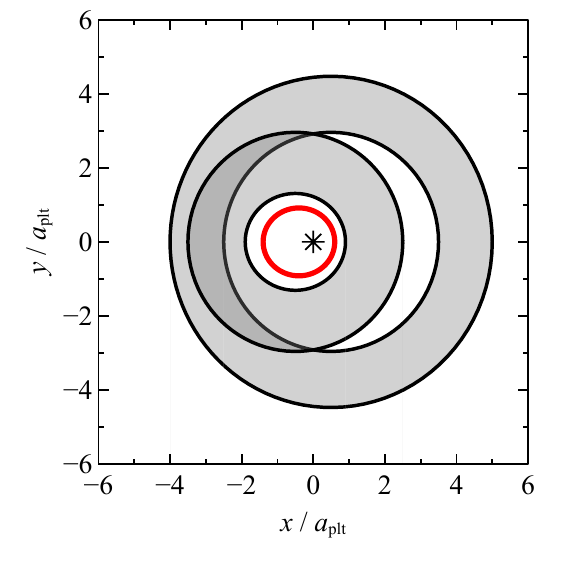}
   \caption{General shape of a debris disc undergoing a secular interaction with a interior, precessing planet. The grey regions are disc material, the red ellipse denotes the planet's orbit and the asterisk shows the star. There are two distinct debris populations; an inner population apsidally aligned with the planet, and an outer one which is antialigned. The inner population may or may not be present depending on the parameters of the system, and likewise for the outer population.}
   \label{fig: general_disc_shape}
\end{figure}

Equation \ref{eq: A} shows that the particle's precession rate $A$ is set by the perturber's mass and the semimajor axes of the particle and perturber, all of which are constants in the secular problem. On Figure \ref{fig: A_vs_a} we plot $A$ as a function of test particle semimajor axis for an example system, which contains a precessing planet. Generally, $A \rightarrow 0$ if the test particle's semimajor axis is very small or very large, and $A \rightarrow \infty$ if the semimajor axes of the particle and planet are similar. Hence if $\dot{\varpi}_{\rm plt} \neq 0$ then there will always be a region in particle semimajor axis space where $A > \dot{\varpi}_{\rm plt}$, and two regions where $A < \dot{\varpi}_{\rm plt}$. In this case a sufficiently broad disc of test particles will cover both the $A > \dot{\varpi}_{\rm plt}$ and $A < \dot{\varpi}_{\rm plt}$ regimes. Figure \ref{fig: general_disc_shape} shows a schematic of the resulting debris structure, for particles external to the planet. The result is a superposition of the debris structures on Figure \ref{fig: ew_evo}, with the innermost particles in the $A > \dot{\varpi}_{\rm plt}$ regime and the outermost in the $A < \dot{\varpi}_{\rm plt}$ regime. There exists a crescent-shaped region devoid of debris, and also a location where $A > \dot{\varpi}_{\rm plt}$ and $A < \dot{\varpi}_{\rm plt}$ particles overlap. However this overlap does not necessarily correspond to debris overdensity, and it is unlikely to be a site of increased dust production; the collision velocities between particles are actually greater within the inner (aligned) disc than those in this inner-outer disc overlap region.

One mechanism which may cause planet precession is the secular effect of the debris. If the planet is much more massive than the total disc mass, the precession rate of the former will be slower than that of the debris, unless the disc extends very close to or far from the star. Hence for all but the broadest discs, if the planet is much more massive than the disc then the result will be an eccentric debris structure apsidally aligned with the planet's orbit. This was the case investigated in \citet{Pearce14}. Alternatively if the planet mass is comparable to that of the disc (ignoring planet evolution for now), the planet will likely precess more rapidly than the outermost debris. Thus the farthest debris will assume an eccentric structure antialigned with the planet's pericentre. Whether the innermost debris also forms this structure, or forms a structure apsidally aligned with the planet (as on Figure \ref{fig: general_disc_shape}), depends on the parameters of the system; a sufficiently eccentric planet will eject all particles with similar semimajor axes, leaving only distant debris which may be slowly precessing. In this case only the antialigned debris structure would be formed. This also leads to a counter-intuitive result; a planet of comparable mass to the disc will clear a larger region of debris than a much more massive planet. This is because particle orbits antialign with that of a low mass (i.e. rapidly precessing) planet, and are therefore more likely to be ejected than if under the influence of a more massive planet (which their orbits would align with).

Thus far we have only considered the secular evolution of debris which does not intersect the planet's orbit. Particles with orbits crossing that of the planet will eventually be scattered unless in a mean-motion resonance, however secular evolution may still occur before then. \citet{Beust14} simulated the evolution of debris under the influence of an eccentric planet, when the planet's orbit crosses that of the debris. They showed that the secular interaction still drives up particle eccentricities as before. However their orbits do not preferentially align or antialign with that of the planet, but rather initially orientate themselves such that for much of the time their pericentres are misaligned with the planet's by $\sim 70 ^\circ$. We observed similar evolution of particles on planet-crossing orbits in our simulations using high mass planets \citep{Pearce14}, suggesting that this behaviour arises because the orbits intersect. That we are now concerned with comparable planet and disc masses does not make a difference; the particles affected by this mechanism have semimajor axes similar to the planet's, and therefore still precess faster than the latter even if the planet is rapidly precessing. Thus in addition to the long-term secular structures described above, particles crossing the planet's orbit will have their eccentricities driven up and their orbits misaligned with the planet's by $\sim 70^\circ$, before eventually being scattered.


\subsection{The effect of scattering on debris}
\label{sec: scattering}

\noindent Material which regularly crosses the planet's orbit (again, if not in a mean-motion resonance) may be scattered by the planet. A particle's post-scattering orbit may differ significantly from its pre-scattering orbit, but both must pass through the scattering point. Hence a planet which scatters material at all points around its orbit will form an overdensity of debris tracing its orbit, caused by the overlapping orbits of scattered particles. This overdensity is often strongest at planet apocentre, where scattering is most efficient; this is because the planet spends more time around apocentre than at other points in its orbit, and the relative velocity between the planet and (non-scattered) disc particles is smallest here.

Objects repeatedly scattered by the planet will eventually leave the system or collide with other bodies. In the meantime, scattered objects may attain very large semimajor axes and eccentricities, and hence their orbits could extend far beyond the initial outer edge of the disc. Inclinations are typically excited less than eccentricities in repeated scattering encounters \citep{Ida92}, so scattered material will form a broad disc superimposed on the secular debris structure discussed in Section \ref{sec: secular_interactions}.

The surface density $\Sigma$ of such a scattered disc will follow an $r^{-3.5}$ profile. This is an empirical result which is observed in all of our simulations regardless of parameters (as well as those of \citealt{Duncan87}), but it can also be obtained using the following semi-analytic method. According to the model of \citet{Yabushita80}, particles repeatedly scattered by a planet will diffuse in semimajor axis space, such that the number of particles $n$ with $x$ in the range $x \rightarrow x + {\rm d}x$ (where $x \equiv 1/a$) at time $t$ is given by

\begin{equation}
n(x,\tau) = \frac{4}{x \tau} \exp\left[-\frac{8}{\tau} \left(1+ \sqrt{x/x_0} \right)\right] I_2 \left[ \frac{16}{\tau} \sqrt[4]{x/x_0}\right].
\label{eq: yabushita_n}
\end{equation}

\noindent Here $x_0$ is the initial value of $x$, $I_2$ is the modified Bessel function, and $\tau \equiv t / t_D(x_0)$ where $t_D(x)$ is the diffusion timescale:

\begin{equation}
t_D(x) \equiv 0.01 T_{\rm plt} \sqrt{a_{\rm plt} x} \left(\frac{M_{\rm plt}}{M_*}\right)^{-2},
\label{eq: t_D}
\end{equation}

\noindent where $T_{\rm plt}$ is the orbital period of the perturbing planet. A reasonable approximation is that scattered particles diffuse in $x$ whilst their pericentre distance and inclination remain constant \citep{Duncan87}. Hence we may build a simple model of the system, whereby debris particles initially on circular orbits with semimajor axes equal to $a_{\rm plt}$ diffuse in $x$, whilst their pericentres remain at $a_{\rm plt}$. We calculate the distribution of $x$ values at time $t$, where $t \gg t_D(x_0)$, using Equation \ref{eq: yabushita_n}. We then create a virtual scattered disc consisting of a large number of particles, with semimajor axes drawn from the above distribution and pericentres equal to $a_{\rm plt}$. For each orbit we calculate the instantaneous radial distance $r$ of the particle at a randomised mean anomaly, and calculate the surface density profile resulting from the summation of these $r$ values for all particles. Regardless of the parameters used (planet mass, semimajor axis and stellar mass) this profile always goes as $r^{-3.5}$. Hence an $r^{-3.5}$ profile appears to be a natural consequence of scattering, and a population of scattered material could potentially be identified from such a slope.

 
\subsection{Planet evolution}
\label{sec: planet_evolution}

\noindent Unlike when the planet is much more massive than the disc, if the two are of comparable mass 
then the planet's orbit may undergo significant evolution. It was noted in Section \ref{sec: secular_interactions} that secular perturbations from the disc cause the planet's orbit to precess, and its orbital plane will also evolve if the planet and disc planes are initially misaligned (although no significant plane evolution will occur if the two are roughly coplanar at the start of the interaction). However the planet's eccentricity may evolve significantly, through planet-particle scattering and secular interactions with the disc. These mechanisms individually affect eccentricity in different ways, so overall the planet's eccentricity behaviour combines two effects. Secular perturbations from the disc will cause the planet's eccentricity to increase and decrease periodically, whilst scattering damps the eccentricity and will circularise the orbit (given enough scattering events). Hence the planet's eccentricity will undergo a long-term decline, with additional oscillatory behaviour in the meantime. If the planet scatters sufficient material before circularisation then there may be too little debris remaining to continue the damping process; in this case, the planet's eccentricity may not tend to zero but to some higher value.

Scattering will also change the planet's semimajor axis. For a single planet scattering debris, the lack of interior planets to remove material scattered inwards means than particles may only leave the system through collisions or ejection. The former mechanism will be rare as the star and planet pose small targets, hence the eventual location of scattered material is likely exterior to its initial orbit. The planet will lose energy to counter this increase in particle energy, hence its semimajor axis will tend to decrease. In section 5.3 of \citet{Pearce14} we derived a theoretical upper limit on this semimajor axis change, and showed that a planet cannot undergo significant migration if much more massive than the disc except for a contrived set of circumstances. The same arguments still apply even when the planet and disc are of comparable mass; in the context of the parameters in equations 14-16 of \citet{Pearce14}, we require $\Gamma \sim 1$ for significant migration, which is unlikely for broad discs. Hence scattering is unlikely to cause any significant change in planet semimajor axis. Recalling that secular interactions also have no effect on this quantity, we conclude that the semimajor axis of an eccentric planet interacting with a comparable-mass debris disc is unlikely to evolve significantly.


\subsection{General numerical simulations}
\label{sec: simulations}

\noindent We now present $n$-body simulations of an eccentric planet interacting with a comparable-mass coplanar debris disc, to demonstrate the physical effects described in Sections \ref{sec: secular_interactions} to \ref{sec: planet_evolution}. We ran almost 100 $n$-body simulations using the Mercury 6.2 integrator \citep{Chambers99}, covering a broad region of parameter space. The general simulation setup is as follows. A planet of mass $M_{\rm plt}$ orbits a star of mass $M_*$, with an initial semimajor axis $a_{\rm plt}$ and eccentricity $e_{\rm plt}$. The planet's pericentre is typically of order 1-10 au; this is roughly the location of the water snow line for solar-type stars, and hence the region where giant planets may be expected to form. The planets have initial eccentricities ranging from 0.1 to 0.9. We fix $M_* = 1 M_\odot$; changing this parameter affects the timescales in the interaction, but not the nature of the evolution.

The star also hosts a debris disc exterior to the planet's pericentre, of mass $M_{\rm disc}$ (where $M_{\rm plt} = M_{\rm disc}$), composed of $N$ equal-mass particles. The disc midplane lies in the planet's orbital plane. We consider discs with initial inner and outer radii ($r_1$ and $r_2$ respectively) of the order of $10-100$ au. The semimajor axes $a$ of disc particles have initial values between $r_1$ and $r_2$, and are distributed such that

\begin{equation}
n(a) \propto a^{1-\gamma},
\label{eq: disc_init_SD}
\end{equation}

\noindent where $\gamma$ is the surface density index. These particles are initially on circular orbits, which are randomised in longitude of ascending node and have inclinations up to an opening angle $I$ with respect to the disc midplane. We use a pre-interaction opening angle of $I = 5^\circ$, that of the classical Kuiper Belt \citep{Bernstein04}, and $\gamma = 1.5$, that of the Minimum Mass Solar Nebula \citep{Hayashi81}, in our simulations. 

We probe disc and planet masses from 0.1 Earth masses (0.1 $M_\oplus$) to 3 Jupiter masses. The discs contain $N = 10^3 - 10^4$ equal-mass debris particles, representing the more massive bodies (i.e. those unaffected by radiation pressure and PR-drag); hence only gravitational forces are included. Each particle exerts a force on the planet and \textit{vice-versa}, but does not perturb other debris. Thus we ignore the self-gravity of the disc; this is discussed in Section \ref{subsec: self_gravity}. Each simulation lasts of the order of $10 - 100$ Myr.

\begin{figure}
  \centering
   \includegraphics[width=8cm]{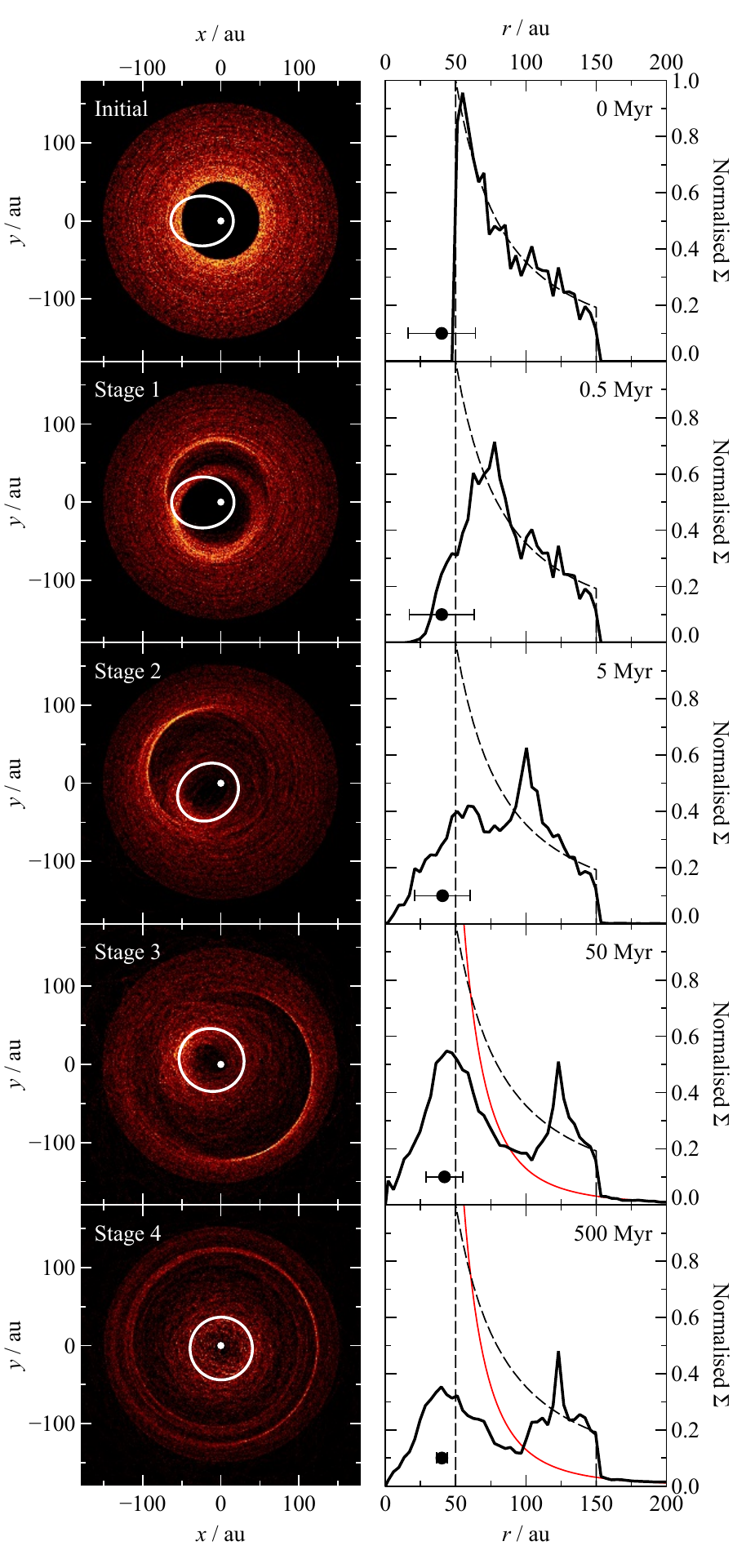}
   \caption{Example $n$-body simulation of an interaction between an eccentric planet and an equal mass, coplanar debris disc, with the simulation parameters described in the text. The left panels show the debris surface density and the planet's orbit (white ellipse), at time zero and then at subsequent evolutionary stages. The right panels show the radially averaged surface density at these times; the thick black lines are the surface density profiles, the dashed lines are the analytic surface density at $t = 0$, and the points show the planet's semimajor axis and pericentre/apocentre distances. The red line on the Stage 3 and 4 surface density plots shows an $r^{-3.5}$ profile, typical of scattered debris, and material beyond 150 au follows this profile. The evolutionary stages are common to all our simulations, and are described in Section \ref{sec: simulations}.}
   \label{fig: position_SD_evo}
\end{figure}

\begin{figure}
  \centering
   \includegraphics[width=7cm]{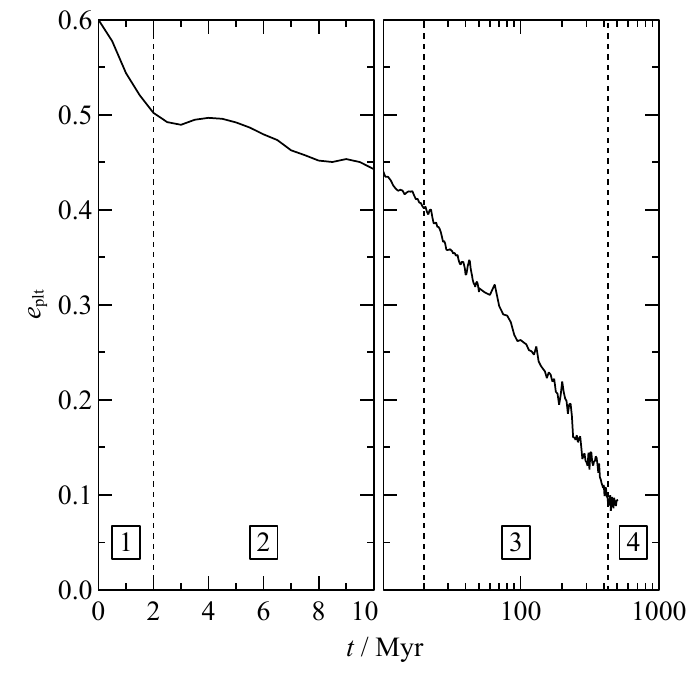}
   \caption{Eccentricity evolution of the planet in the simulation shown on Figure \ref{fig: position_SD_evo}, as an example of the general behaviour observed in all our simulations. At early times (up to $\sim10$ Myr) secular eccentricity oscillations are noticeable, on top of the long-term decline from debris scattering. The dotted lines and numbers in boxes refer to the stages of system evolution, for comparison with Figure \ref{fig: position_SD_evo}.}
   \label{fig: eplt_evo}
\end{figure}

Despite the broad range of parameters tested, the interaction always produced the same qualitative results. We found four distinct evolutionary stages occurring on logarithmic timescales, which we describe below. We also present an example simulation; we show the disc surface density at each evolutionary stage on Figure \ref{fig: position_SD_evo}, and the planet's eccentricity evolution on Figure \ref{fig: eplt_evo}. The example shows a 10 $M_\oplus$ planet interacting with a comparable-mass disc, with $a_{\rm plt} = 40$ au, $e_{\rm plt}=0.6$, $r_1 = 50$ au, $r_2 = 150$ au and $N=10^3$.

\begin{description}
  \item[Stage 1:] The planet begins to scatter material from the inner regions of the disc, depleting the debris surface density inwards of planet apocentre. Non-scattered particles with orbits crossing that of the planet have their eccentricity increased via the planet's secular influence; these orbits are preferentially misaligned by $\pm\sim 70^\circ$ to the planet's orbit, due to the effect described in \citet{Beust14}. Particles beyond the planet's apocentre still have roughly circular orbits, and the similar secular phases of neighbouring particles cause the formation of a spiral-shaped overdensity beyond the planet's orbit \citep{Wyatt05}. The planet's eccentricity undergoes its most rapid decline due to debris scattering, and secular effects may also cause this eccentricity to oscillate. \\

  \item[Stage 2:] All debris initially crossing the planet's orbit has been scattered at least once; an overdensity of scattered material forms along the planet's orbit, and this overdensity is strongest around planet apocentre. A population of scattered material with surface density going as $\sim r^{-3.5}$ begins to form beyond the planet's orbit, extending beyond the initial outer edge of the disc. This population is initially small, and hence is not visible on Figure \ref{fig: position_SD_evo} until the final panel. However a logarithmic surface density plot demonstrates that a $\sim r^{-3.5}$ population has begun forming by the second stage. Material originating just exterior to the planet's orbit may form a coherently eccentric disc apsidally aligned with the planet, depending on the planet's precession rate and eccentricity (see Section \ref{sec: secular_interactions}). The spiral-shaped overdensity continues to develop beyond the planet's orbit. The debris surface density profile hence has two peaks: a broad peak of scattered material stretching between the planet's initial pericentre and apocentre distances, and sharp peak farther out corresponding to the spiral overdensity. \\

  \item[Stage 3:] At least one secular period has elapsed for particles initially orbiting just exterior to the planet's apocentre; some initially stable material has been driven onto eccentric orbits apsidally antialigned with the planet, crossed its orbit and been scattered. Hence the surface density of scattered material tracing the planet's orbit is increased, and a large crescent shaped gap forms in the disc in the direction of planet pericentre. The spiral overdensity exterior to the planet continues to move outwards, as more distant particles are still in secular phase with their neighbours. The planet's eccentricity may by now have reduced significantly. Note that the planet's location corresponds to a region of overdensity in the disc, rather than the region of underdensity (as might na\"{i}vely be expected). \\

  \item[Stage 4:] The planet has scattered all material crossing its orbit, and its eccentricity evolution essentially ceases. Hence the innermost peak of the surface density profile has been reduced or even removed. An overdensity of scattered material may still exist just exterior to the planet's orbit; this material no longer comes close to the planet since the latter's eccentricity decreased, so this debris is now stable. Particles driven to high eccentricities by secular effects early on may now have their eccentricities frozen, as the forcing eccentricity becomes small owing to the decrease in planet eccentricity. If the planet circularisation timescale is much longer than the secular timescale of the outermost particles then surviving non-scattered debris forms a smooth disc apsidally antialigned with the planet; otherwise, the spiral overdensity may still be present in the outer debris and remain there indefinitely.

\end{description}

Generally, whilst the planet's eccentricity and longitude of pericentre evolve significantly throughout the interaction, its other orbital elements remain roughly constant. In the example simulation $a_{\rm plt}$ changes by less than 5 per cent, and $i_{\rm plt}$ never exceeds $0.6^\circ$ (from an initial value of $0^\circ$).

The qualitative results presented in this section are general. The quantitative results will differ for specific systems, but rough scaling rules can be applied. For example, increasing the mass of the planet and disc simultaneously will decrease the interaction timescales, whilst increasing the planet semi-major axis and disc radii will increase timescales. Increasing the planet eccentricity will decrease the circularisation timescale, and moving the disc mass inwards (either through reducing $r_1$ or increasing $\gamma$) makes the planet circularise faster and to a greater degree.

Whilst this paper primarily considers the case where $M_{\rm plt} = M_{\rm disc}$, the results are applicable to the $M_{\rm plt} > M_{\rm disc}$ regime too. Even if the planet were orders of magnitude more massive than the disc, the general secular behaviour is the same as in the equal mass case. A difference between the two mass regimes is that the transition between aligned and antialigned particles occurs farther from the star if the planet is more massive than the disc; this is because the planet would precess more slowly relative to debris than the equal mass case, so the location where particles precess more slowly than the planet is farther from the star (see Figure \ref{fig: A_vs_a}). This is why we did not observe this secular behaviour in \citet{Pearce14}; our discs simply did not extend far enough outwards to probe this regime. The main qualitative evolutionary difference between the equal mass case and that when the planet is much more massive is that the planet will not undergo the same degree of orbital evolution in the latter regime.

An interesting result may occur if $1 \lesssim M_{\rm plt}/M_{\rm disc} \lesssim 10$, whereby particles can change between the aligned and antialigned secular regimes. This occurs because the planet initially precesses rapidly (leading to antialignment of some particle orbits), yet the planet is massive enough to eject a significant fraction of the disc particles. The declining disc mass causes the planet precession to slow, meaning that the precession rate of some particles can ``overtake'' that of the planet. The final result of such an interaction is the formation of a coherently eccentric disc (as in \citealt{Pearce14}), but with a more messy structure due to additional particles which have changed their secular behaviour. We do not wish to comment on the case where $M_{\rm plt} < M_{\rm disc}$, as disc self gravity would be very important in this regime and thus the results of this paper probably do not apply there (see Section \ref{subsec: self_gravity}).

Our results may be used to predict the outcome of an eccentric planet interacting with a coplanar debris disc of the same or greater mass. They may also be used to infer the presence of an unseen perturber from the structure of an imaged debris disc, as we will now demonstrate for HD 107146.


\section{Application to HD 107146}
\label{sec: HD107146 Application}

\noindent HD 107146 is a 80-200 Myr old G2V star, located 27.5 pc from the Sun \citep{van_Leeuwen07, Williams04}. In 2000, IRAS imaging revealed excess infrared emission in the stellar spectrum, indicative of a debris disc \citep{Silverstone00}. As noted in Section \ref{sec: Introduction}, further observations resolved the disc in both infrared emission and scattered light \citep{Ardila04, Williams04, Carpenter05, Corder09, Hughes11}. For an excellent summary of work on HD 107146 up until 2011, see \citet{Ertel11}.

\begin{figure*}
  \centering
   \includegraphics[width=12cm]{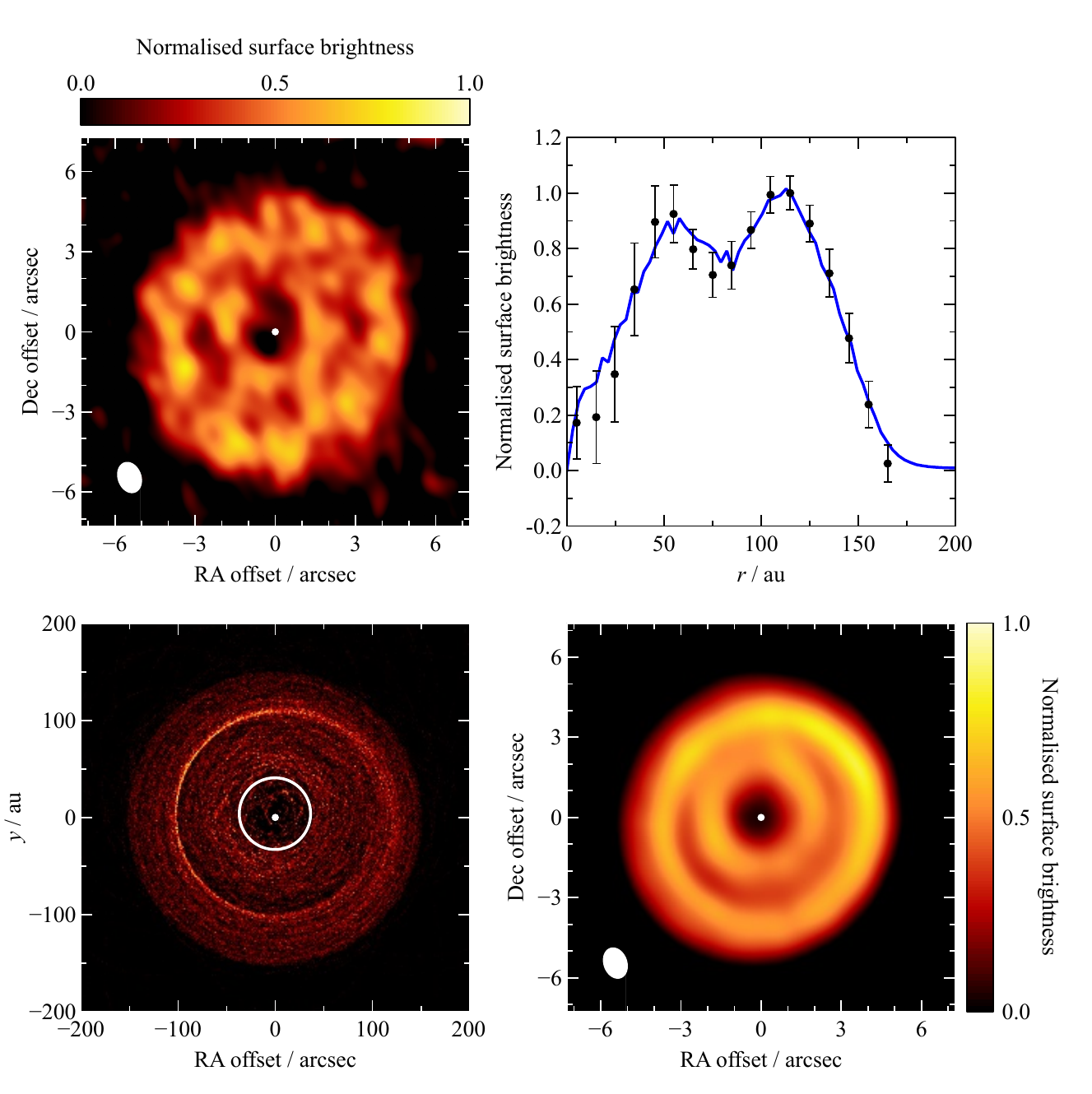}
   \caption{ALMA observations of HD 107146, along with our best-fitting simulation. Top left: ALMA 1.25mm continuum image \citep{Ricci15}. Note that we use a different colour scale to that in the aforementioned paper. The white ellipse represents the beam size and orientation. Top right: points show the normalised, radially-averaged surface brightness profile of the disc as observed by ALMA, measured using elliptical apertures. The solid line is the profile from our best-fitting $n$-body simulation, at the time (19 Myr after the start of the interaction) of the best fit; the two agree with a reduced $\chi^2$ value of 0.4. The simulation parameters and the method used to compare the data and simulation are described in Section \ref{sec: HD107146 Application}. Bottom left: positions of debris particles in the best-fitting $n$-body simulation at 19 Myr. The $x-y$ plane is the initial disc midplane, with planet pericentre initially pointing along the $x$ axis. The orbit of each particle has been populated with 100 points with randomised mean anomalies, to increase the effective number of particles plotted. The white point is the star, and the white ellipse the planet's orbit. Bottom right: simulated ALMA image of the $n$-body disc. The particles have been scaled for emission, the image rotated, and smoothed with a 2D Gaussian representing the ALMA beam (white oval). Compare this to the ALMA observation in the top left, noting that we have not added noise and hence our image is smoother.}
   \label{fig: best_fit_obs_pos_SD}
\end{figure*}

The recent 1.25mm ALMA image reveals the disc at millimetre wavelengths in unprecedented detail \citep{Ricci15}. These data show that the disc spans $30-150$ au from the star and, assuming it is circular, inclined by $21^\circ$ to the sky plane at a position angle (E of N) of $140^\circ$. These observations detected $0.2 {\rm M}_\oplus$ of dust at 1.25mm, and by extrapolating this up to bodies of diameter $D = 1000$ km (with the number of bodies of a given diameter $n(D) \propto D^{-3.6}$) the authors inferred a total disc mass of $100 {\rm M}_\oplus$. The ALMA image and corresponding surface brightness profile are shown on the top two plots of Figure \ref{fig: best_fit_obs_pos_SD}.

It is clear from these plots that the disc has an unusual morphology. The outer regions are brighter than the inner, and the brightness profile decreases with radius before increasing again farther out. Lower-resolution 880$\mu$m SMA observations also show that the surface brightness does not decrease with radius as expected \citep{Hughes11}. These data have been interpreted as the disc's surface density profile either being double-peaked or increasing with radius with a gap at $80$ au \citep{Ricci15}, and these models are indistinguishable at the observation resolution. Possible causes of these profile include embedded Pluto-sized objects inducing collisions between large debris bodies, or perturbations from an unseen planetary companion. The ALMA observations failed to detect any CO gas, suggesting that a dust-gas interaction is not responsible for the disc morphology.

We wish to ascertain whether a past (or ongoing) interaction between the disc and a hypothetical eccentric planet can explain the disc features, and if so, estimate the pre-interaction orbit of the planet as well as its present day location. We assume the planet originated interior to the disc, where some event placed it onto an eccentric orbit; this could have been a planet-planet scattering or merger event, for example \citep{Lin97, Ford08}. We aim to reproduce the 1.25mm ALMA observations with an $n$-body simulation of such an interaction. Millimetre grains are unaffected by radiation pressure and PR-drag, so should act as tracers of the parent debris bodies (those most important for the dynamics of the system). Hence the ALMA data is well-suited to modelling with purely gravitational $n$-body simulations.


\subsection{Simulation setup}
\label{subsec: HD107146 sims}

In addition to the simulations described in Section \ref{sec: simulations}, we ran a further $\sim 150$ simulations specifically aimed at reproducing the HD 107146 debris disc. We describe their setup now. HD 107146 is a G2V star, so we fix its mass at $1 {\rm M}_\odot$. We also fix the disc mass to the $100 {\rm M}_\oplus$ value of \citet{Ricci15}. However this still leaves nine physical variables: the planet's mass, its initial semimajor axis, eccentricity, inclination and argument of pericentre, and the disc's initial inner and outer radii, opening angle and surface density profile. Computational limitations prevent us from exploring this whole parameter space, so we make several assumptions about the pre-interaction system to reduce the number of variables.

We again fix the pre-interaction disc opening angle at $I = 5^\circ$ and $\gamma = 1.5$, and again assume the planet initially orbits in the disc midplane. These assumptions leave five physical parameters: $M_{\rm plt}$, $a_{\rm plt}$, $e_{\rm plt}$, $r_1$ and $r_2$. However we can use physical reasoning to fix a further two of these. Firstly, the disc of HD 107146 appears to be roughly axisymmetric. In \citet{Pearce14} we showed that if $M_{\rm plt} \gg M_{\rm disc}$, the planet's eccentricity will not be significantly damped by the disc, and external debris will form a coherently eccentric disc aligned with the planet's orbit. Conversely, in Sections \ref{sec: planet_evolution} and \ref{sec: simulations} we showed that if $M_{\rm plt} \sim M_{\rm disc}$ then the planet's eccentricity will be significantly damped, and hence the outer edge of the disc will remain roughly circular. Thus if HD 107146's disc is interacting with a reasonably eccentric planet (or did so in the past) then $M_{\rm plt} \sim M_{\rm disc}$, so we fix the planet mass to be $100 {\rm M}_\oplus$ in our simulations. This means that the outer edge of the disc will be largely unchanged by the interaction, so we fix $r_2 = 151$ au (which best fits the data in the outermost regions). Again, we use $N = 10^3 - 10^4$ equal-mass debris particles to simulate the disc, and omit disc self-gravity (see Section \ref{subsec: self_gravity}).

We are thus left with three free parameters: the initial values of $a_{\rm plt}$ and $e_{\rm plt}$, and the initial inner disc radius $r_1$. These three are somewhat degenerate, so we cannot fix any at a single value. Instead, we run simulations with $a_{\rm plt} = 20$, 30, 40 and 50 au (noting that the inner peak of the observed surface density profile is at 50 au, and that the planet's initial semimajor axis is typically interior to this peak in our general simulations), and for each $a_{\rm plt}$ we run simulations with various values of $e_{\rm plt}$ and $r_1$. We disfavour simulations where the planet's initial pericentre is within 3 Hill radii of the disc inner edge, or external to this location; in these cases the disc would be unstable before the interaction started. Once our simulations are complete we compare them to the observations, using the method described below.


\subsection{Constructing simulated observations}
\label{subsec: HD107146 convert to obs}

Throughout each simulation we compare the instantaneous distribution of debris to that observed by \citet{Ricci15}. This requires the simulated debris to be converted into an image and surface brightness profile as would be observed by ALMA, for which we use the following method. Firstly, we populate each particle's orbit with 100 points at randomised mean anomalies, to increase the effective number of particles simulated. We then scale for emission by weighting each point by a black body; a point at radial distance $r$ from the star is weighted to have a luminosity $L$, where

\begin{equation}
L(r) \propto B_\nu (\lambda, T).
\label{eq: weight_plancks_law}
\end{equation}

\noindent Here $B_\nu (\lambda, T)$ is the spectral radiance of a body of temperature $T$ at a wavelength $\lambda$, given by Planck's law:

\begin{equation}
B_\nu (\lambda, T) \propto \left[\exp\left(\frac{h c}{\lambda k_{\rm B}T}\right) - 1\right]^{-1},
\label{eq: plancks_law}
\end{equation}

\noindent where $h$ is the Planck constant, $c$ is the speed of light and $k_{\rm B}$ is the Boltzmann constant. The temperature of the body is determined by the flux it receives from the star (again assuming black body behaviour), hence

\begin{equation}
T = \left(\frac{L_*}{4 \pi \sigma} \right)^{1/4} r^{-1/2}
\label{eq: T}
\end{equation}

\noindent where $L_*$ is the star's luminosity and $\sigma$ is the Stefan-Boltzmann constant. For HD 107146, we use $L_* = L_\odot$ and $\lambda = 1.25$mm, that of the ALMA observations. For these parameters, $B_\nu (\lambda, T)$ roughly scales as $r^{-1/2}$. For this analysis we have assumed the disc is optically thin; this is valid here since, whilst the disc is massive, it covers a broad region. Whilst the optical depth could be high if the disc were extremely thin, even a moderate $5^\circ$ opening angle would be large enough that we do not have to consider flux attenuation here.

To produce images for comparison with the ALMA observations, we rotate the simulated (emission scaled) disc to an inclination of $21^\circ$ and a position angle of $143^\circ$. We then convolve our image with a two-dimensional Gaussian to simulate the ALMA point spread function (PSF); this Gaussian has a standard deviation along its major axis of 13.4 au, along its minor axis of 9.8 au, and its major axis has a position angle of $19.8^\circ$. We also calculate the radially averaged surface brightness profile of the simulated disc, using elliptical apertures on the simulated image as in \citet{Ricci15}. We may then compare our simulations to the ALMA observations of HD 107146.


\subsection{Fitting the HD 107146 disc}
\label{subsec: fitting HD107146}

\begin{figure*}
  \centering
   \includegraphics[width=16cm]{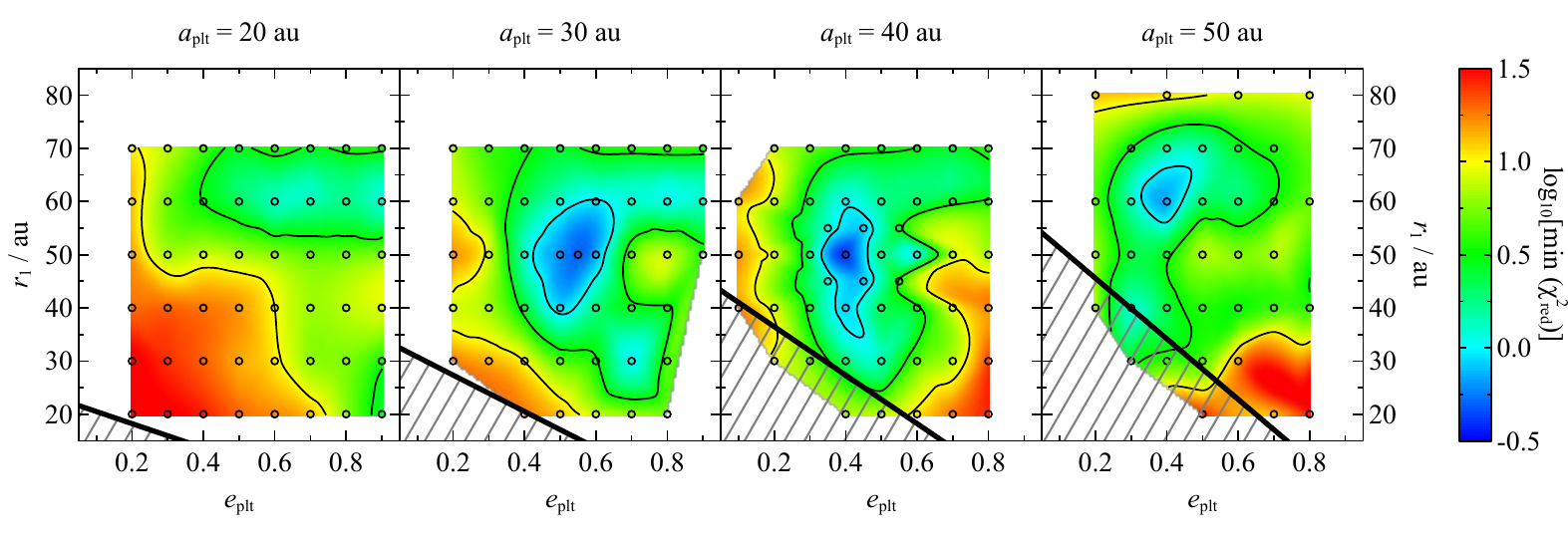}
   \caption{Reduced $\chi^2$ of our simulated radially-averaged surface density profiles compared to that of HD 107146, at the time in each simulation when this parameter is minimised. We varied the initial inner disc radius $r_1$ and the initial planet eccentricity $e_{\rm plt}$ for four initial planet semimajor axes $a_{\rm plt}$, fixing all other parameters as described in the text. The points show our simulations, with the colourmap and contours interpolated between them. Contours show $\log_{10}[\min(\chi^2_{\rm red})] = 0$, 0.5 and 1 respectively. The hatched regions show an unphysical area of parameter space, where the planet's initial pericentre is closer than three Hill radii to the disc inner edge (see Section \ref{subsec: HD107146 sims}).}
   \label{fig: chi2_maps}
\end{figure*}

\noindent We identified the simulations which best replicate the HD 107146 disc using a $\chi^2$ analysis. At 100 time intervals throughout each simulation we calculated the $\chi^2$ value comparing the observed radial surface brightness profile with the simulated profile at this time (found using the method in Section \ref{subsec: HD107146 convert to obs}). On Figure \ref{fig: chi2_maps} we plot the $a_{\rm plt}$, $e_{\rm plt}$, $r_1$ parameter space tested in our simulations, and colour each point by $\min(\chi^2_{\rm red})$ (the minimum value of reduced $\chi^2$, that is the minimum value of $\chi^2$ attained during that simulation, divided by the number of degrees of freedom). If the data are independent, a reduced $\chi^2$ of order 1 means that the obs1ervations are consistent with the model, and the smaller the value, the better the fit (although values much smaller than 1 imply the data is overfitted). Here the observed surface brightness profile points are correlated with each other, so little should be inferred from the exact value of $\min(\chi^2_{\rm red})$ itself; however this value does allow a comparison between simulations, to identify that which best reproduces the HD 107146 disc.

Figure \ref{fig: chi2_maps} shows that there are several regions of tested parameter space which produce discs consistent with ALMA observations. The best fit is attained using a planet with initial semimajor axis $a_{\rm plt} = 40$ au and eccentricity $e_{\rm plt} = 0.4$, interacting with a disc with initial inner radius $r_1 = 50$ au. For this case the simulated surface brightness profile is most similar to the ALMA observations 19 Myr after the start of the interaction, and we plot the simulated ALMA image and surface brightness profile at this time on the lower two plots of Figure \ref{fig: best_fit_obs_pos_SD}. These well reproduce the observations; the surface brightness profile yields a $\min(\chi^2_{\rm red})$ value of 0.4, and the simulated image resembles the ALMA observation by eye. Note that globular structures in the observed image are probably noise, which has not been accounted for in the simulated image and hence the latter appears smoother than the observation. This best fit occurs when the simulated system is at stage 3 or 4 in its evolution (as described in Section \ref{sec: simulations}); the planet has removed most of the material crossing its orbit, and its orbital evolution has essentially stalled. By this point the planet's eccentricity has decreased from 0.4 to 0.05, whilst its semimajor axis (initially 40 au) has only reduced by 4 au. The simulation first reaches $\chi^2_{\rm red} \sim 1$ at 10 Myr, and this parameter remains less than 1 until the end of the simulation (at 30 Myr); hence the simulation also provides a good fit to the observations over a long time interval.

However the best-fitting simulation is not unique in reproducing the observations. A well-defined $\chi^2$ minimum also exists for a planet semimajor axis of 30 au, centred on $e_{\rm plt} = 0.55$ and $r_1 = 50$ au, and this is almost as good as the best-fitting 40 au solution ($\min(\chi^2_{\rm red}) = 0.5$). Again, the planet in the $a_{\rm plt} = 30$ au simulation undergoes minimal semimajor axis evolution whilst its eccentricity is significantly reduced, and the simulation fits best once it has evolved to stage 3 or 4 and resembles the observations for a long time. Conversely, we find that planets with initial semimajor axes of 20 and 50 au do not reproduce the observed disc well. Also note that our well-fitting simulations have the disc's initial inner edge exterior to its present day value, and material has since been scattered inwards by the planet. In conclusion, a planet with an initial semimajor axis of 30 or 40 au and an eccentricity of 0.4-0.5, interacting with a comparable-mass debris disc with initial inner edge at 50 au, can well reproduce the disc of HD 107146. At present, the planet is most likely on a roughly circular orbit at 30-40 au.


\section{Discussion}
\label{sec: discussion}

\noindent We have examined the general interaction between an eccentric planet and a coplanar, comparable-mass debris disc, and applied our results to HD 107146 in an attempt to explain its unusual disc. In this section we discuss a possible limitation of our work: the omission of disc self-gravity. We also discuss the timescale of the HD 107146 interaction. Finally, we examine the implications of this paper for planet searches, both for general systems with debris discs and also for HD 107146.


\subsection{Disc self-gravity}
\label{subsec: self_gravity}

\noindent Our simulated debris particles exert a force on the planet (and \textit{vice-versa}), but do not interact with each other; hence we do not include disc self-gravity in our simulations. This omission dramatically increases computational efficiency, allowing us to run several hundred simulations for this paper. However whilst self-gravity does not affect the interaction outcome if the planet is much more massive than the disc (as in \citealt{Pearce14}), if the two are of comparable mass then this effect could become important.
 
Debris in a self-gravitating disc would undergo additional secular and scattering evolution from the influence of other disc particles. Secular interactions work over large distances on timescales scaling inversely with the object masses, whilst scattering works over short distances on timescales going as the inverse-square of the masses (see equations 17 and 18 in \citealt{Pearce14}). Therefore given the small debris particle masses, the major effect of self-gravity is likely to be on the secular evolution of the disc.

We investigate the possible secular effect of self-gravity by analytically calculating the precession rate of a test particle embedded in a disc. This gives us a feel for how the initial disc in our best-fitting HD 107146 simulation would evolve due to self-gravity alone (in the absence of any planetary perturbations), and allows us to compare the magnitude of the self-gravity effect to that of the planet. 

To calculate the precession rate, we consider a test particle at position ($R$, $\phi$, $z$) in cylindrical coordinates, which experiences a force from a 2 dimensional, axisymmetric disc in the $z=0$ plane. Equation 2-146 in \citet{Binney87} gives the radial acceleration of the particle due to the disc as

\begin{multline}
F_{\rm r}(R) = -\frac{G}{R^{3/2}} \int_{r_1}^{r_2} \bigg[K(k) - \frac{1}{4} \frac{k^2}{1-k^2} \\ \times \left(\frac{R'}{R} - \frac{R}{R'} + \frac{z^2}{R'R}\right) E(k) \bigg] k \Sigma(R') \sqrt{R'} {\rm d}R',
\label{eq: Fr}
\end{multline}

\noindent where $R'$ is the radial location of a point in the disc,

\begin{equation}
k^2 \equiv \frac{4 R R'}{(R + R')^2 + z^2},
\label{eq: k}
\end{equation}

\noindent and $K(k)$ and $E(k)$ are the complete elliptical integrals of the first and second kind respectively. We wish to consider a particle in the disc midplane, i.e. $z=0$. However \citet{Binney87} note that Equation \ref{eq: Fr} has an unphysical singularity at $R = R'$ if $z=0$, because $k=1$ here and so the $K(k)$ and $(1-k^2)^{-1}$ terms become undefined. This issue can be resolved by setting $0 < z \ll R$, so we use $z=10^{-4}$ au in our evaluation. A particle in the midplane experiences no vertical acceleration, and its tangential acceleration is also zero because the disc is axisymmetric. Hence the self-gravity of an axisymmetric disc exerts only a radial force on a disc particle. The precession rate of a particle at true anomaly $f$ moving in a Keplerian potential and perturbed by an additional radial force $F_{\rm r}$ is

\begin{equation}
\dot{\omega} = -\frac{1}{e}\sqrt{\frac{a(1-e^2)}{\mu}}F_{\rm r} \cos f
\label{eq: wdot}
\end{equation}

\begin{figure}
  \centering
   \includegraphics[width=7cm]{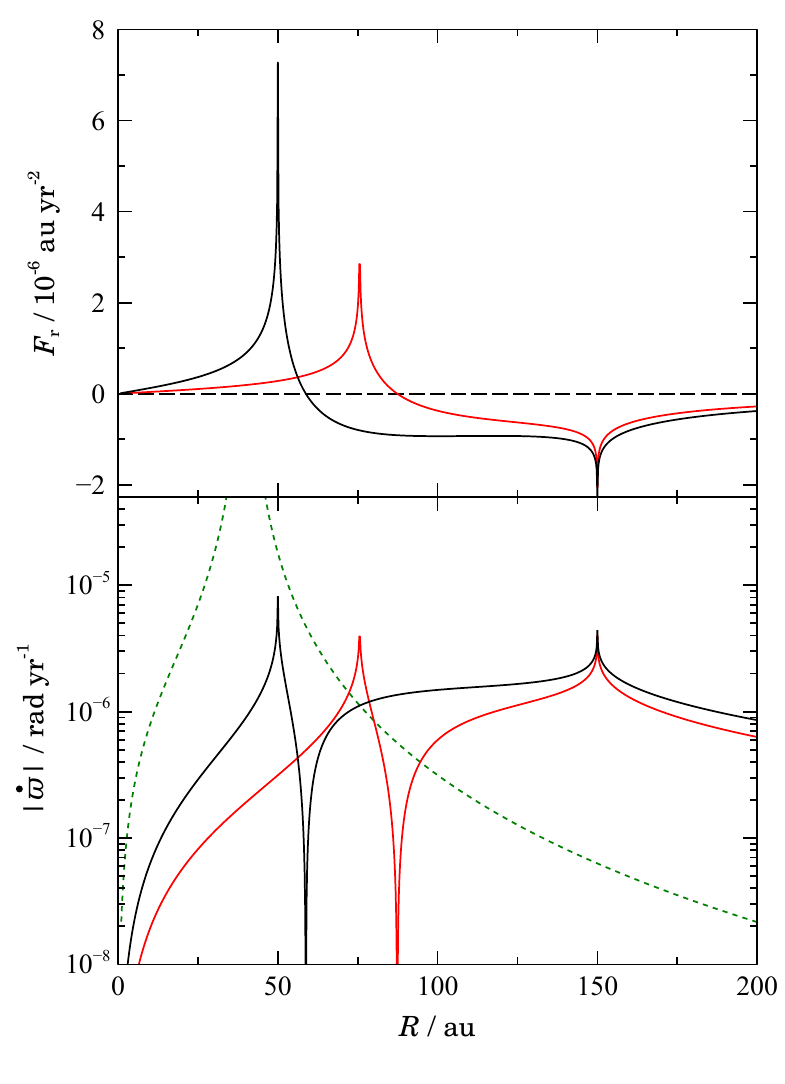}
   \caption{The evolution of a test particle under the influence of a massive axisymmetric disc. Top plot: the radial force imparted by the disc on a coplanar particle at radius $R$, from Equation \ref{eq: Fr}. The black line shows a 100 $M_\oplus$ disc with an $r^{-1.5}$ surface density profile, with inner and outer radii of 50 and 150 au respectively. These are the initial disc parameters in the best-fitting HD 107146 simulation. The red line shows the same disc but with all mass inwards of 75 au removed, representing the truncation of the disc by the planet. Bottom plot: the magnitudes of the resulting particle precession rates. The green dotted line shows the precession rate due to the secular influence of a 100 $M_\oplus$ planet with $a_{\rm plt} = 40$au (Equation \ref{eq: A}). The plot shows that the planet dominates particle evolution in the inner regions of the disc, whilst disc self-gravity may be more important in the outer regions.}
   \label{fig: self_grav_disc}
\end{figure}

\noindent (section 2.9 of \citealt{Murray_Dermott99}), and we average this over the orbital period $T$:

\begin{equation}
\langle\dot{\omega}\rangle \equiv \frac{1}{T}\int_0^T \dot{\omega} {\rm d}t \approx \frac{1}{2\pi a^2 \sqrt{1-e^2}}\int_0^{2\pi} \dot{\omega} r^2 {\rm d}f, 
\label{eq: wdot_av}
\end{equation}

\noindent where we have assumed that $a$ and $e$ are constant over one orbital period. Finally, if $e \ll 1$ then $F_{\rm r}$ will not vary significantly over the particle's orbit. In this case

\begin{equation}
\langle\dot{\omega}\rangle \approx \sqrt{\frac{a}{\mu}} F_{\rm r}(a),
\label{eq: wdot_av_fin}
\end{equation}

\noindent and similar analyses for semimajor axis and eccentricity yield $\dot{a} \approx \dot{e} \approx 0$. We evaluated Equation \ref{eq: Fr} numerically for a disc with the initial parameters of that in our best-fitting HD 107146 simulation. The force, and the resulting precession rate, are shown as functions of radius by the black lines on Figure \ref{fig: self_grav_disc}. 

The plot shows that the precession rate of particles inwards of 75 au is still dominated by the planet, even when disc self-gravity is considered. In the $n$-body simulation (without self-gravity), the planet drives up the eccentricities of these particles and scatters the majority of them, with the remainder forming an eccentric disc aligned with the planet's orbit. Hence this would still occur with the inclusion of self-gravity. However the disc's gravity may initially dominate beyond this region; in the $n$-body simulation, particles beyond 80 au preferentially antialign with the planet's orbit, so debris out to 100 au is removed. Self-gravity would effectively cause these particles' precession rates to be uncorrelated with the planet's evolution, preventing both preferential antialignment and also significant eccentricity excitation. So were disc self-gravity included, the depletion of the disc in the best-fitting HD 107146 simulation would initially extend out to 75 au rather than 100 au. This would not fit the observational data.

However we have not yet considered the evolution of the disc self-gravity. Particles inwards of 75 au would be depleted even with self-gravity, and this would change the disc potential. The red lines on Figure \ref{fig: self_grav_disc} show a disc with the same parameters as discussed above, but with all mass inwards of 75 au removed. Now the planet is still influential out to about 95 au, so much of this debris may still eventually undergo scattering. Beyond this region the disc self-gravity will always dominate, although in the simulation these particles were not significantly perturbed by the planet anyway. Hence the inclusion of self-gravity will not affect the overall simulation results in the outermost regions.

Our analysis suggests that, for our best-fitting HD 107146 simulation, the inclusion of disc self-gravity would not qualitatively affect the resultant disc structure interior to 75 au and exterior to 95 au. In the region between these radii, the potential effect of self-gravity is unclear. This region might not undergo the same level of depletion as in the simulations, and the spiral structure visible in Figure \ref{fig: best_fit_obs_pos_SD} might not be present. Hence our observational fit might not be as good as that on Figure \ref{fig: best_fit_obs_pos_SD}. More generally, depending on the simulation parameters, self-gravity may affect our predicted outcomes for the interaction investigated in this paper. The main effect of self-gravity would probably be the reduction of debris depletion in the region immediately interior to the outermost peak of the disc. However we stress that a more sophisticated self-gravity analysis is required to fully explore its potential effect, which is beyond the scope of this paper.


\subsection{HD 107146 interaction timescale}
\label{subsec: timescales}

\noindent Our best-fitting HD 107146 simulations all reproduce the observed disc $\sim10$ Myr after the start of the interaction, compared to the 100 Myr age of the star. These two timescales are compatible, but scenarios in which they are comparable would be preferable. The interaction timescales are set by the disc (and hence planet) mass and, since the disc mass derived from observations is uncertain \citep{Ricci15}, there is scope to change this in our simulations. The secular interactions between the planet and disc are the dominant effects in the simulations, and Equations \ref{eq: e_forced} - \ref{eq: laplace_coeff} show the secular precession rate to scale linearly with mass whilst the forcing eccentricity is independent of mass. Equation \ref{eq: Fr} shows that the effect of disc self-gravity also scales linearly with disc mass, so scaling both $M_{\rm plt}$ and $M_{\rm disc}$ simultaneously will not change the importance of self-gravity relative to planetary perturbations. Hence changing the disc and planet masses should affect the secular interaction timescales, but not the nature of this interaction. Reducing the masses will make the planet less efficient at ejecting debris, but seeing as the main effects of the interaction are secular in nature, this should not affect the outcome too much. Hence if we reduce the disc and planet masses in our simulations by an order of magnitude (so $M_{\rm disc} \sim 10 {\rm M}_\oplus)$, then roughly the same interaction will occur over a timescale comparable with the stellar lifetime. Hence whilst our interaction timescales are by no means incompatible with the system age, if we assume that this interaction is responsible for the observed disc structure then our results might suggest the disc mass is closer to $10 {\rm M}_\oplus$ than $100 {\rm M}_\oplus$. Alternatively the planet may have only recently been placed on an eccentric orbit, and we happen to have observed the system at this stage in its evolution.


\subsection{Implications for planet searches}
\label{subsec: planet_searches}

Our findings have interesting implications for the inference of unseen planets from debris disc features, both generally and for HD 107146. An important result is that the planets in this interaction generally circularise with little change in semimajor axis, having ejected much of the debris interior to their final orbital distance. Hence the eventual location of the planet is typically near the inner edge of the disc, on an orbit traced by a debris overdensity, beyond which lies a gap followed by another overdense ring. This configuration would not otherwise be expected; having observed a double-peaked debris disc, the na\"{i}ve assumption would be that any perturbing planet lies in the underdense region between the two peaks. Hence future planet finding missions should not be discouraged if no planets are found in a debris disc gap; indeed, the absence of planets in this region may hint at a violent dynamical history, and could motivate the search for planets near the inner edge of the disc instead.

If our hypothesis on the evolution of the HD 107146 system is correct, then a $\sim 100 M_\oplus$ planet currently orbits near the inner edge of the disc, with a semimajor axis of $30 - 40$ au and an eccentricity of $\sim 0.1$. This planet originated interior to the disc; assuming it was scattered out of its original location by another body then, based on its initial pericentre, a second companion with mass at least equal to that of the scattered planet exists at $\gtrsim10-25$ au from the star. Companions of less than 10 Jupiter masses ($3000 M_\oplus$) have not been ruled out anywhere in the system by imaging \citep{Apai08}, so this scenario is possible and could be tested with deeper planet searches. Furthermore, our simulations suggest that the outermost debris peak actually forms a thin spiral, rather than a continuous ring. This structure would be detectable in observations with $\sim 3$ times the resolution of the ALMA image, and such a detection would favour our hypothesis on the history of the system (although with the caveat that the disc self-gravity would also have an effect, and may partially or completely wash-out this spiral). Such a resolution may well be possible with current instrumentation.

Another potential application of this work is to HD 92945; this system may also harbour a double-peaked debris disc \citep{Golimowski11}, so our results could be used to invoke a perturbing planet in that system too. However the HD 92945 disc was imaged in scattered light, so the emitting dust would be affected by radiation forces. Hence it is unclear without more detailed analysis whether the more massive debris also follows this double-peaked profile (as in HD 107146), or whether the observed morphology is a consequence of non-gravitational forces on small dust.


\section{Conclusions}
\label{sec: conclusions}

\noindent Broad, double-ringed debris discs could potentially have evolved to their present state under the influence of an eccentric, comparable-mass planet. We investigate this interaction in general, and show that it follows four distinct stages on logarithmic timescales. A key result is that planet precession may cause distant debris orbits to \textit{anti-}align with that of the planet, whilst the innermost debris orbits align with the planet's. This results in distinct inner and outer debris regions with a gap or depletion between them, akin to the double-peaked debris structures potentially observed in HD 107146 and HD 92945. It also produces the counter-intuitive result that a low-mass planet may clear a larger region of debris than a higher-mass body. In general the planet undergoes a rapid eccentricity decrease whilst its semimajor axis remains constant; thus if the planet initially scattered off another body then the two would quickly decouple, so our results still hold in the presence of additional massive planets (providing the eccentricity damping is fast enough).

We then modelled the HD 107146 system in detail, confirming that the debris disc's unusual morphology can be well explained by this interaction. If an unseen eccentric planet did sculpt debris into the structure seen today, then this hypothetical planet initially had pericentre in the inner regions of the system and apocentre within the disc itself; based on our best-fitting model, the planet is currently on a low-eccentricity orbit 30-40 au from the star. This is below the companion detection thresholds of current observations of the system, but could potentially be found by future imaging projects.


\section{Acknowledgements}
\noindent We thank Luca Ricci for allowing us the use of his ALMA image, and Mher Kazandjian for discussions concerning the modelling of disc self-gravity. We also thank Herv\'{e} Beust for his very constructive and helpful review. TDP acknowledges the support of an STFC studentship, and MCW is grateful for support from the European Union through ERC grant number 279973.


\bibliographystyle{mn2e}
\bibliography{bib_HD107146}


\label{lastpage}

\end{document}